\documentclass[%
 aip,
 amsmath,amssymb,
 reprint,%
]{revtex4-1}

\usepackage{graphicx}
\usepackage{dcolumn}
\usepackage{bm}

\usepackage[utf8]{inputenc}
\usepackage[T1]{fontenc}
\usepackage{mathptmx}
\usepackage{etoolbox}
\usepackage{soul}
\usepackage{color}

\makeatletter
\def\@email#1#2{%
 \endgroup
 \patchcmd{\titleblock@produce}
  {\frontmatter@RRAPformat}
  {\frontmatter@RRAPformat{\produce@RRAP{*#1\href{mailto:#2}{#2}}}\frontmatter@RRAPformat}
  {}{}
}%
\makeatother
\begin{document}

\preprint{AIP/123-QED}
\title[Sample title]{Confinement and shear effects on the rotational diffusion of a minimal virus-inspired colloidal particle }
\author{Karen Gonzales-Flores}
\author{Ramón Castañeda-Priego}
 \author{Francisco Alarcón}\email{paco@fisica.ugto.mx}
\affiliation{ 
División de Ciencias e Ingenierías, Universidad de Guanajuato, Loma del Bosque 103, 37150, León, Mexico 
}%


\date{\today}

\begin{abstract}
The rotational diffusion of a rigid spherical body decorated with dimers in an explicit fluid environment is reported. This model acts as a simplified representation of an enveloped virus bearing peplomers and immersed in a coarse-grained fluid. Using dissipative particle dynamics, we explicitly study the hydrodynamic effects on the rotational diffusion of this virus-inspired particle subjected to oscillatory shear flow and confined between two solid-like surfaces. Since the rotational response is affected by the type of imposed flow, we first characterize the oscillatory shear by identifying distinct flow regimes in terms of the Péclet number, $Pe$. Our findings suggest that, under confinement, the rotational diffusivity is primarily controlled by the amplitude of the oscillatory flow, whereas the contribution of the number of peplomers remains secondary. This behavior is likely related to their small size and dimeric structure, which only weakly perturb the hydrodynamic response. For high and lower $Pe$, the rotational diffusion coefficient, $D_{r}$, tends to decrease as the number of peplomers ($N_{s}$) increases, while at intermediate values of $Pe$, the interplay between oscillatory forcing and thermal fluctuations prevents the appearance of a clear trend between $D_{r}$ and $N_{s}$. Our results provide a general picture of how in confined environments the interplay between fluid flow and thermal fluctuations can affect the rotational diffusion of spiked particles, thereby helping to explain how fluid conditions can modify the alignment of peplomers with their potential targets.
\end{abstract}

\maketitle

\begin{quotation}
\end{quotation}

\section{\label{sec:level1}INTRODUCTION\lowercase{} }
Soft matter systems, such as colloidal dispersions, liquid crystals, and polymeric solutions, among others, can generally be described at time and length scales much larger than the atomistic one \cite{Weik}. 
Studying these systems at full atomistic resolution is demanding, especially when trying to accurately represent their hydrodynamic behavior \cite{MOEENDARBARY,Zhou}.
Therefore, technically speaking, a purely microscopic approach, where all degrees of freedom on a wide variety of time and length scales coexist, becomes intractable to describe the dynamics of such systems on measurable time scales \cite{EspanolWarren,Groot}.  
 
Consequently, several continuum theories have been adopted, including Navier-Stokes-Fourier hydrodynamics for complex fluids \cite{EspanolWarren}. These continuum approximations can be applied at various scales, as they are treated as coarse-grained particles or fields that represent groups of atoms or molecules \cite{Landau, Groot, EspanolWarren}. 
Within this approach, the continuum limit corresponds to a volume element that contains a sufficiently large number of particles to reproduce the thermodynamic behavior of the whole system. This allows a physical description on macroscopic length scales through nonlinear differential equations, which are typically solved numerically \cite{EspanolWarren}.
Alternatively, numerical simulations also help to understand experimentally inaccessible processes \cite{Shillcock}. 
Furthermore, in soft matter systems, the characteristic energy scale is of the order of thermal energy, which makes it necessary to include fluctuating terms in both continuum theories and numerical simulations \cite{Groot, Chen}. Theoretical and computational frameworks that connect microscopic (atomistic) views of matter with macroscopic (continuum) descriptions are referred to as mesoscopic models.

Mesoscopic models are capable of reproducing the main dynamical features of complex systems on long time and length scales \cite{Malevanets}. Depending on the characteristics of the problem,  different techniques have been proposed to simulate hydrodynamic interactions at the mesoscale, such as grid-based methods, e.g., lattice Boltzmann (LB), particle-based methods, e.g., smoothed particle hydrodynamics (SPH), multiparticle collision dynamics (MPCD), and dissipative particle dynamics (DPD) \cite{MOEENDARBARY, Groot, Amati, Zantop}. 
A key characteristic of the latter approach is that it explicitly accounts for thermal fluctuations \cite{Groot, Espanol}. 

DPD allows the representation of the fluid as soft particles and correctly includes hydrodynamic interactions \cite{Curk,Espanol,Yawei}. This simulation method enables the modeling of solvents in complex fluids to examine their transport properties. For example, it can be used to study the translational and rotational diffusion of colloidal suspensions composed of rigid rods, spheres, and other bodies. Thus, colloidal suspensions described by simple shapes or geometries act as effective analogs for many real-world systems, such as polymers, proteins, and lipids. \cite{ZZhang,Chen,Yawei}.

In this work, we introduce a fluid model consisting of a rigid spherical body decorated with dimers that serves as a minimal representation of an enveloped virus with peplomers and simulate its dynamics in an explicit environment using the DPD method. By imposing an oscillatory shear flow on the fluid, we investigate the rotational diffusion of the virus model confined between two walls. The rationale for studying the rotational behavior of a decorated virus model arises from the observation that enveloped viruses depend on rotational motion to orient their surface peplomers toward the surrounding environment, a mechanism that can affect how these peplomers become optimally aligned with receptors on host cells \cite{Kanso, Boschi, Bakhshandeh}.
Furthermore, the mobility and rotational dynamics of such viruses are influenced by the properties of the host medium, which can modulate viral motion and, in turn, may impact infection-related processes \cite{Coccia, Shen, Vahey, Wallace}. Therefore, DPD will allow us to quantitatively examine how the combined effects of fluid flow, confinement, and thermal fluctuations collectively affect the rotational behavior of spike-decorated particles.

One prominent example of the class of virus-like systems that we focus on, which can be described at the mesoscopic level using our virus model, is the Severe Acute Respiratory Syndrome Coronavirus 2 (SARS-CoV-2) suspended in saliva \cite{Dbouk} and confined in the respiratory tract. This virus is transmitted by inhalation of respiratory droplets containing the virus, leading to influenza-like symptoms \cite{Sharma, Seminara}. Although the disease mainly affects the respiratory system, it can also cause damage to many vital organs \cite{Laue, Kopanska}. This enveloped virus is characterized by the presence of spike glycoproteins, also known as peplomers, distributed on its surface \cite{Dbouk, Sharma}, making it a representative example of a spike-decorated particle immersed in a complex and confined fluid environment. Although the present work does not specifically aim to model SARS-CoV-2, this example illustrates a wider range of biological systems that inspire investigations into the rotational behavior of spike-covered particles in complex flows, because viruses bearing peplomers on their surfaces depend on these structural features to diffuse and reorient effectively in particular environments \cite{Coccia, Shen}.
In addition, the morphology of enveloped viruses plays a crucial role in evading the first line of defense of the body \cite{Bustamante}, which is designed to immobilize pathogens as they traverse the mucosal barriers of the host environment\cite{Wallace, Vahey}. 
For example, by performing site-specific fluorescent labeling and super-resolution microscopy experiments, Vahey et al. \cite{Vahey} demonstrated that the organization and dynamics of surface proteins in the influenza A virus (IAV) facilitate penetration of the host mucus.

In general, viruses of this type exhibit variations in their physical, electrostatic, structural, and hydrophobic properties, as well as in the dynamic distribution of peplomers on their surfaces, resulting from conformational alterations \cite{Cotten, Lu, Triveri, Wales}. These alterations can be the result of mutations that occur during the viral replication cycle, producing variability in surface characteristics and peplomer arrangement \cite{Bakhshandeh}. This variability complicates the identification of universal transport mechanisms and motivates the use of simplified physical models to isolate the role of structural features. Additionally, the number of peplomers has been shown to vary even among nearly genetically identical viruses,  as reported by Laue et al. \cite{Laue}.

The impact of viral architecture on diffusion was recently analyzed from first principles \cite{Kanso} using a rigid bead-rod model, where the spatial arrangement of peplomers on the surface of the capsid is obtained by applying an energy minimization criterion. The authors demonstrated that the rotational diffusivity of the virus decreases monotonically with increasing peplomer concentration on its surface. 
Subsequently, modifications were introduced to both the envelope \cite{Kanso22} and the peplomers \cite{Kanso21}. These findings showed that both the elliptical shape of the viral envelope and the triangular structure of the coronavirus peplomer decrease the rotational diffusivity of the virus. In a related study, Moreno et al. \cite{Moreno} investigated the motion of various enveloped viruses by solving the Stokes equations within the framework of the rigid multi-blob (RMB) model. They examined both the translational and rotational diffusivities of viruses whose surfaces are decorated with spikes.
Their results demonstrated that the number and distribution of the spikes, as well as the morphology, significantly influence diffusion, consistent with previous findings.
More recently, mesoscopic simulations employing smoothed dissipative particle dynamics (SDPD) have been carried out \cite{Moreno25}, where peplomers and viral envelope were treated as separate rigid bodies to capture both the tilting of the spikes and their diffusion over the envelope surface.
Furthermore, the dynamics and spatial distribution of the peplomers were characterized, revealing a clustering behavior consistent with experimental observations from DNA-PAINT super-resolution imaging of HIV-like particles \cite{Moreno25}.

The previous studies showed that the density, shape, spatial arrangement, and mobility of peplomers are closely related to the diffusive behavior of the virus. Nevertheless, the inclusion of each contribution in any virus-like (and more realistic) model represents a challenging and demanding (computational) task. Thus, simplified or coarse-grained models allow us to understand the role of each degree of freedom on the transport properties of viruses under specific thermodynamic conditions. Therefore, in this contribution, we focus on a minimal viral structural model with controlled peplomer distributions and populations. Peplomers are modeled as dimers to introduce a minimal degree of surface anisotropy and to emphasize how the interplay between oscillatory flow and confinement shapes the rotational dynamics of an enveloped virus. Our simplified approach allows us to examine the rotational diffusivity of the virus in terms of an oscillatory shear flow in competition with thermal fluctuations. To this end, we systematically vary the Péclet number $(Pe)$, which is defined as the ratio of the characteristic shear rate to diffusive transport. Moreover, because viruses are typically found in confined environments such as the respiratory tract, we represent the spike-covered virus as a rigid particle coated with dimers, suspended in a coarse-grained DPD fluid confined between two solid surfaces. As described and analyzed in this work, these conditions give rise to hydrodynamic behavior that is distinct from that found under bulk conditions.

\section{Dissipative Particle Dynamics and Viral Suspension Model}\label{DPD-II}
\subsection{DPD method}
DPD is a mesoscopic, particle-based simulation methodology in which each particle represents a coarse-grained segment of the fluid under investigation. The dynamics of these particles is governed by Newton's second law. Consequently, the equation of motion for each particle $i$ can be expressed as,
\begin{eqnarray}
 \bar{F}_{i}=\frac{d\bar{v}_{i}}{dt},\quad  \bar{v}_{i} = \frac{d\bar{r}_{i}}{dt},
\end{eqnarray}
where the particle mass is set to unity for simplicity. As a result, the net force exerted on each particle $i$ is directly proportional to its acceleration \cite{Groot} and can be expressed as a sum of three pairwise forces \cite{Espanol}, namely, the conservative force $\bar{F}^{C}_{ij}$, the dissipative force $\bar{F}^{D}_{ij}$, and the random force $\bar{F}^{R}_{ij}$. These forces are summed over all the $i^{th}$-particle's neighbors as
\begin{equation}
 \bar{F}_{i}=\sum_{j\neq i}{\bar{F}^{C}_{ij}+\bar{F}^{D}_{ij}+\bar{F}^{R}_{ij}}, 
 \label{newton}
\end{equation}
which are given by
\begin{equation} \label{eq1}
\begin{split}
\bar{F}^{C}_{ij} & =  a_{ij}\omega^{C}(r_{ij})\hat{r}_{ij}, \\
  \bar{F}_{ij}^{D} & = -\gamma \omega^{D}(r_{ij})(\hat{r}_{ij}\cdot(\bar{v}_{i}-\bar{v}_{j}))\hat{r}_{ij}, \\
  \bar{F}_{ij}^{R} & = \sigma \omega^{R}(r_{ij})\vartheta  _{ij}\hat{r}_{ij}.
\end{split}
\end{equation}
The parameter $a_{ij}$ represents the maximum repulsive force between the particles $i$ and $j$, while $\gamma$ and $\sigma$ are constants that define the strength of the dissipative and random forces, respectively \cite{Espanol, Gharibvand},  $r_{ij} = |\bar{r}_{i} - \bar{r}_{j}|$ is the distance between the particles $i$ and $j$, and $\hat{r}_{ij} = (\bar{r}_{i} - \bar{r}_{j})/ r_{ij}$ is the normalized direction between the two centers of the particles. $\bar{v}_{i}$ and $\bar{v}_{j}$ are the velocities of the particles $i$ and $j$, respectively.

The conservative force is expressed as a short-range and soft repulsion along the center-to-center direction, since the force is defined up to $r_{ij} < r_c$ and 0 otherwise, additionally the linear weight function is defined as $\omega^{C}(r_{ij})=(1-r_{ij}/r_{c})$; where $r_{c}$ is chosen as the unit of length \cite{Groot, Salib}.
The dissipative force represents a drag force in the direction of relative motion between the particle $i$ and $j$ and is modulated by the weight function $\omega^{D}(r_{ij})$ \cite{Groot, Liu2}.
The random force is characterized by the weight function $\omega^{R}(r_{ij})$, and the term $\vartheta_{ij}$ is a Gaussian white noise with stochastic properties
$<\vartheta_{ij}>=0$ and $<\vartheta _{ij}(t)\vartheta _{kl}(t')>= (\delta_{ik}\delta_{jl}+\delta_{il}\delta_{jk})\delta(t-t')$, where $\delta(t-t')$ is the Dirac delta function \cite{Mai}. Furthermore, the condition $\vartheta_{ij} = \vartheta_{ji}$ is imposed to guarantee the conservation of momentum \cite{Espanol}. 

By writing Newton's second law associated with the DPD forces, it results in a set of Langevin-like equations, which can be rewritten as the Fokker-Planck equation for the positions and momenta of all probability distributions of the particles \cite{EspanolWarren}.
When the thermodynamic system reaches thermal equilibrium, the steady-state solution of the Fokker-Planck equation corresponds to the Gibbs canonical ensemble. Furthermore, fulfillment of the equilibrium state can be guaranteed by imposing the following conditions:
\begin{equation}
   \omega _{R}(r_{ij})=\omega_{D}^{1/2}(r_{ij}),
   \label{termo1}
\end{equation} 
\begin{equation}
    \sigma =(2k_{B}T\gamma )^{1/2}.
    \label{termo}
\end{equation}
Eq. (\ref{termo1}) states that the weight functions of the dissipative and random forces are coupled, and Eq. (\ref{termo}) relates the dissipative force to the random force through the fluctuation–dissipation theorem. 
The conservation of momentum, together with the conditions imposed by Eqs. (\ref{termo1}) and (\ref{termo}), constitutes a set of physical constraints necessary to ensure the correct recovery of hydrodynamic behavior\cite{Nikunen, Espanol}.

\subsection{Fluid model}  

The model fluid environment consists of a generic coarse-grained DPD fluid confined between two parallel walls composed of DPD particles. The positions of the fluid particles are initially randomly generated within a volume whose dimensions are $30 \times 17 \times 30$ (in units of $r_{c}^{3}$), and their velocities follow a Gaussian distribution.    

\begin{figure}[h]
\centering
  \includegraphics[height=4.6cm]{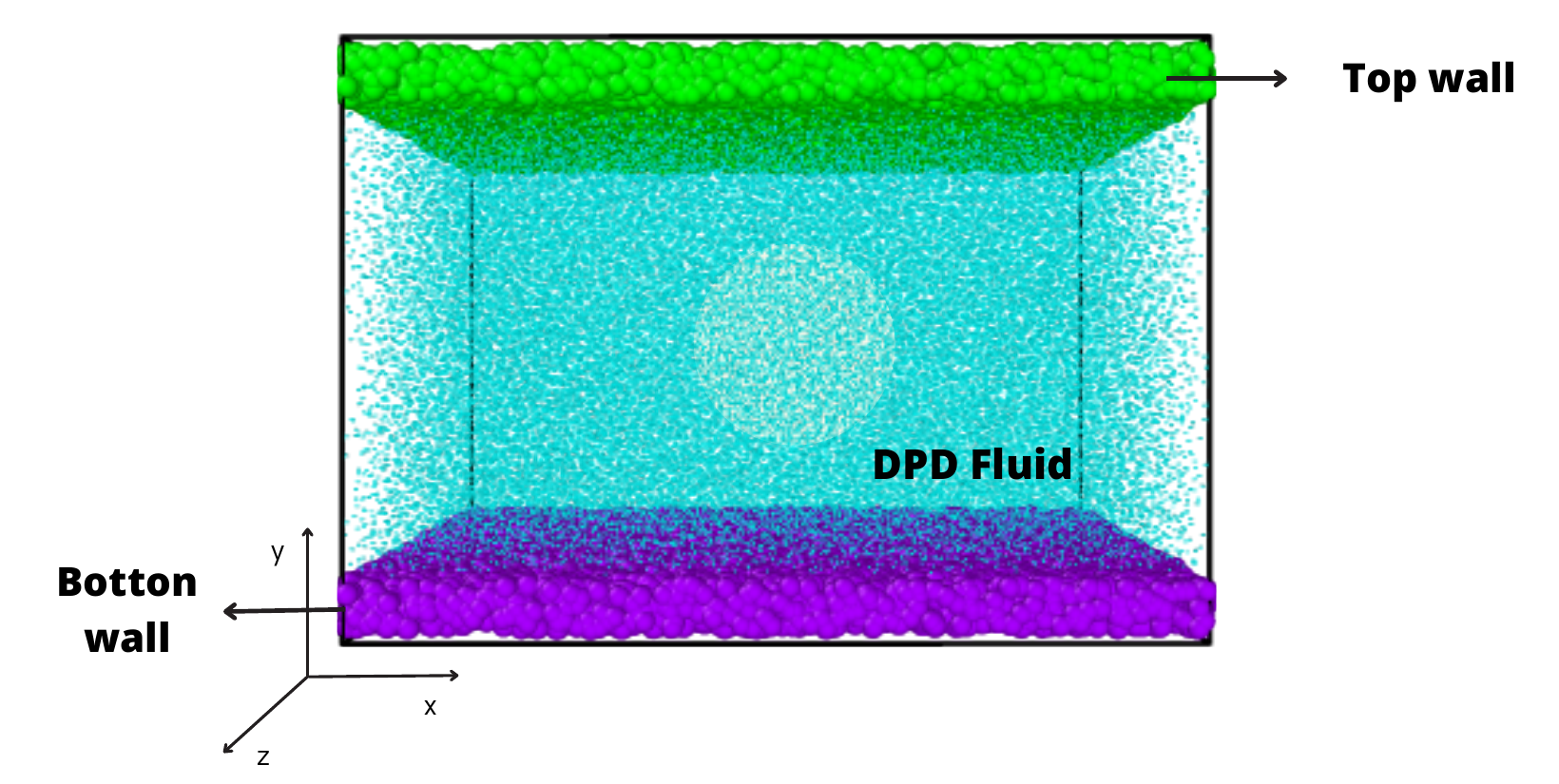}
  \caption{Snapshot of the simulation model, which consists of a DPD fluid between two parallel walls, where periodic boundary conditions are applied in the $x-z$ directions and wall boundary conditions are applied in the $y$ direction. Blue particles represent the fluid, green and purple particles correspond to the top and bottom walls, respectively.}
  \label{sistema}
\end{figure}

Fluid particles do not pass through the walls and their spatial distribution is illustrated in Fig. \ref{sistema} (blue particles). To set up this host fluid model, a spherical region with a radius larger than $4.5 r_{c} $ is left free of fluid particles in the center of the simulation box (see Fig. \ref{sistema}). The rigid spherical body decorated with dimers or the virus model is subsequently placed within this region. This approach ensures that the fluid particles are not initially located within the viral capside.

To preserve the compressibility condition of the host fluid, its density is set to $\rho_{f}=3$ within the simulation domain; a total of $56,700$ fluid particles are included in the computational box. The conservative force between the fluid particles is characterized by a parameter $a_{ff}=75$ (in units of $k_{B}T/r_{c}$) \cite{Groot,Melchor,Wijmans}, where the label $f$ stands for fluid. This specific value is selected to ensure stable fluid behavior and reproduce the expected thermodynamic properties of a DPD fluid under the chosen simulation conditions \cite{Groot}.

Periodic boundary conditions are applied along the $x$ and $z$ directions, while the fluid is confined between two parallel walls in the $y$ direction. The interactions between the fluid and the wall particles are characterized by a repulsion parameter  $a_{fw}=75$, where the subscript $w$ denotes the wall particles.
The value $a_{fw}=a_{ff}$ is chosen to represent hydrophilic walls.

\subsection{Walls model}

The walls are represented by frozen DPD particles. The centers of the wall particles are randomly assigned throughout the $x-z$ simulation plane and within a $1.5r_c$ gap at both boundaries of the simulation box in the $y$ direction. The initial velocities of the wall particles are set to zero ($v_{x}=v_{y}=v_{z}=0$). 
Each wall contains $24,670$ particles, a number carefully chosen to achieve a sufficiently dense packing and thereby prevent fluid particles from penetrating the wall structure. This setup ensures that particles belonging to the walls remain confined within the boundaries of the simulation domain. Technically speaking, these conditions are chosen to ensure the structural stability of the walls, prevent their deformation or collapse throughout the simulation, and maintain the integrity of the confinement boundaries under fluid pressure. Both walls are modeled as waterproof and hydrophilic surfaces. A visual representation of the whole environment considered here is provided in Fig. \ref{sistema}.

The wall, composed of randomly distributed DPD particles, introduces surface roughness and provides a microscopic implementation of the non-slip boundary condition. Momentum-dissipative interactions between the wall and fluid particles slow the fluid near the boundary, enforcing its velocity to match that of the wall. This approach allows for an effective simulation of viscous drag at the interface, in contrast to crystalline wall structures, which can produce, slip conditions, artificial layering or periodicity effects \cite{Pivkin05}. 

\subsection{Virus model}\label{virus-model}

With the DPD technique, we construct rigid spherical bodies from smaller constituent particles, which can represent either a protein or a lipid segment of the viral envelope. In this virus-inspired model, a total number of $819$ DPD envelope particles are positioned in concentric circles, with every particle located at the same distance from the virus center, forming the surface of the virion.
\begin{figure}[h]
\centering
  \includegraphics[height=5cm]{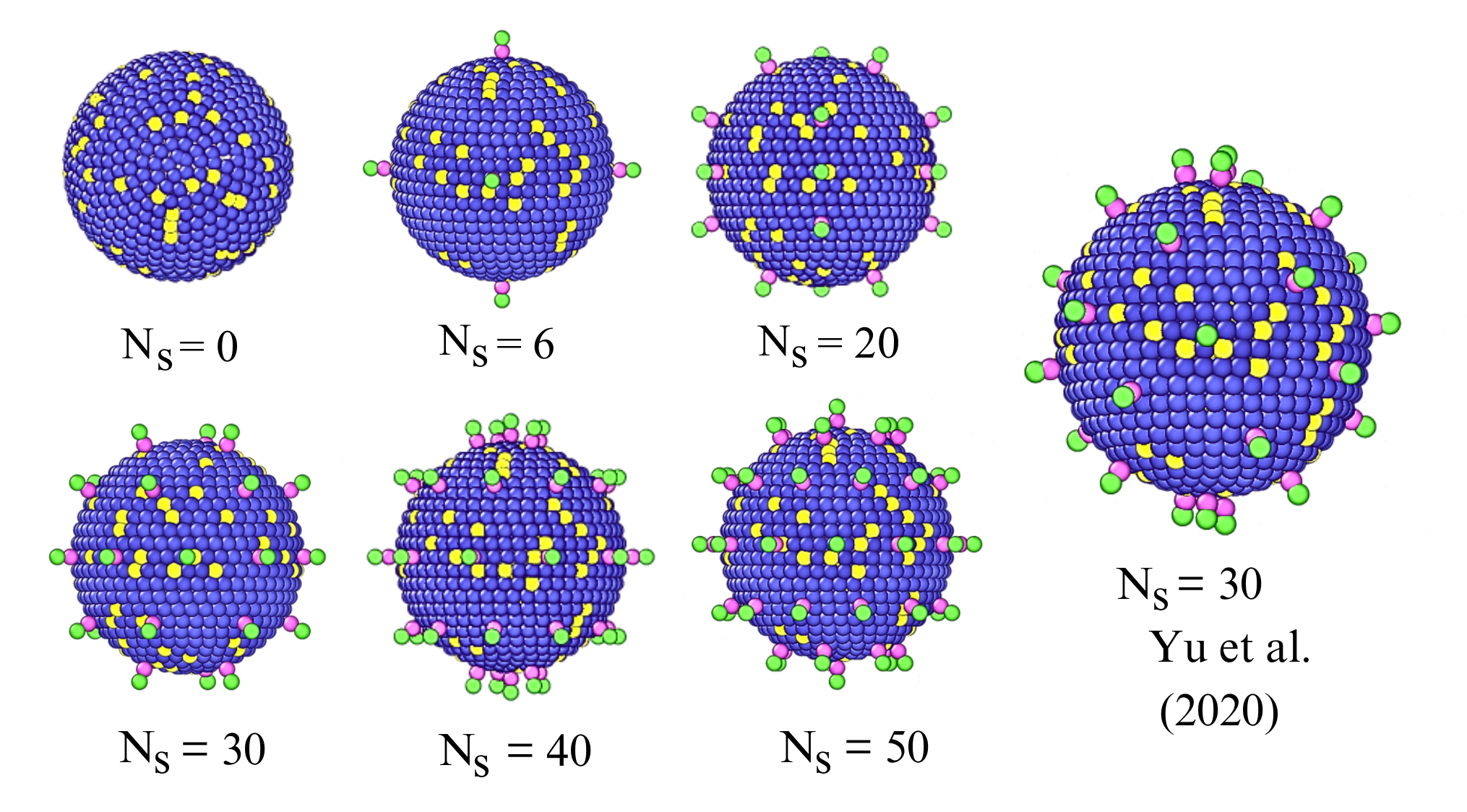}
  \caption{The virus model is composed of DPD particles, where each virus is represented as a rigid structure with a variable distribution of peplomers on its surface.  The number of peplomers $N_{s}$ ranges from 20 to 50 to achieve a homogeneous distribution on the virus surface. The second virus model with $N_{s}=30$ is based on the work of Yu et al.\cite{Yu}. The case without peplomers ($N_{s}=0$) is also considered to provide a meaningful comparison with the virus containing peplomers.}
  \label{detallado}
\end{figure}

We also consider different conformational arrangements for the distribution of peplomers \cite{Laue}. Each peplomer is explicitly modeled as a protrusion from the rigid viral surface, composed of two fixed DPD particles. In general, the peplomers are arranged roughly evenly over the envelope, so that the spacing between adjacent spikes is almost the same throughout. An additional configuration comprising $30$ peplomers is derived from the structural data of Yu et al., where the peplomer locations correspond to experimentally identified positions (see Fig. \ref{detallado}) \cite{Yu}.

As illustrated in Fig. \ref{detallado}, the model includes four types of DPD particles, inspired by the structural components of the SARS-CoV-2 \cite{Yu}. These include: the small envelope protein $E$ (yellow particle), which generally maintains a conserved structure consisting of a short hydrophilic amino-terminal region followed by a large hydrophobic domain \cite{Kuo}; the membrane protein $M$ and the lipid bilayer (blue particles), which are the most abundant components and, for simplicity, are represented by a single DPD particle type \cite{Alharbi}; and the spike glycoprotein, divided into two subunits—$S_{1}$ (green particle), responsible for binding to the host cell receptor, and $S_{2}$ (pink particle), which contains hydrophobic amino-acid-rich domains \cite{Mansbach}.

The dimensions of the virus model are selected to match the characteristic sizes of enveloped viruses, including SARS-CoV and SARS-CoV-2, which typically have diameters of $ \sim 100 \hspace*{0.1cm} nm$ \cite{Neuman, Laue}. These reference values ensure that the DPD simulations maintain physical relevance.
In this work, the radius of the envelope structure is set to $4.5 r_c$, which corresponds to a DPD length unit of $r_c = 11.1$ nm.
The geometric height of the peplomer is $1.5r_c$; however, one-third of this length overlaps with the DPD surface particles of the envelope structure. Therefore, the peplomers have an effective radius of approximately $0.5r_c$. We consider a DPD particle for each subunit, $S_{1}$ and $S_{2}$. 

%
\begin{table}
\caption{\label{table} Repulsion parameter $a_{ij}$ for the $5$ types of DPD particles corresponding to $M, E, S_{1}$ and $S_{2}$ (see text for details). }
\begin{ruledtabular}
\begin{tabular}{ccddd}
$a_{ij}$&$f$&\mbox{$M/E$}&\mbox{$S_{1}$}&\mbox{$S_{2}$}\\
\hline
$f$&$75$&\mbox{$80$}&\mbox{$75$}&\mbox{$90$}\\
$W$&$75$&\mbox{$75$}&\mbox{$75$}&\mbox{$75$}\\
\end{tabular}
\end{ruledtabular}
\end{table}
The interaction affinities between the viral constituents and the fluid particles ($f$), as well as the wall particles ($W$), are summarized in Table \ref{table}. This Table describes the conservative force repulsion parameters for the four types of DPD particles used in the virus model with the fluid and wall particles. Since only one enveloped virus is considered in the host fluid, interactions between viruses are not considered. The particles $S_2$, $M$, and $E$ correspond to partially hydrophobic regions and are thus assigned higher repulsion parameters relative to the fluid \cite{Kuo,Alharbi,Mansbach}.

Finally, the virus is modeled as a rigid independent body.
At each time step, the total force and torque on the virus are calculated as the sum of the forces and torques acting on its constituent particles. The positions, velocities, and orientations of all particles are then updated so that the virus translates and rotates as a single entity \cite{lammps}.

\subsection{Characterization of the hydrodynamic field}

The hydrodynamic field is essential for describing the behavior of the DPD fluid across all Péclet numbers examined. In addition, the diffusivity of the virus model is strongly influenced by the hydrodynamical characteristics of the host fluid. Consequently, a detailed account of the methodology used to determine the hydrodynamic field is necessary. For this purpose, the velocities of all fluid particles are tracked during the last stages of the simulation to determine fluid properties, such as velocity profiles.

To compute the hydrodynamic field, the simulation domain is discretized along the $y$-axis, corresponding to the direction perpendicular to the walls. This axis is divided into equally spaced bin slabs with a volume of $L_xL_z \Delta y$, where $\Delta y=1 r_c$, and $L_x$ and $L_z$ are the box sides along the $x$ and $z$ directions, respectively.  Within each bin, the instantaneous velocities of the fluid particles are averaged to calculate the local mean velocity. By systematically repeating this process for all bins along the $y$-direction, a velocity profile is obtained, providing insight into the flow structure as a function of inter-wall separation for each value of $Pe$.

\subsection{Rotational characterization of the virus\label{TwoF}}

\begin{figure}[h]
\centering
  \includegraphics[height=9.3cm]{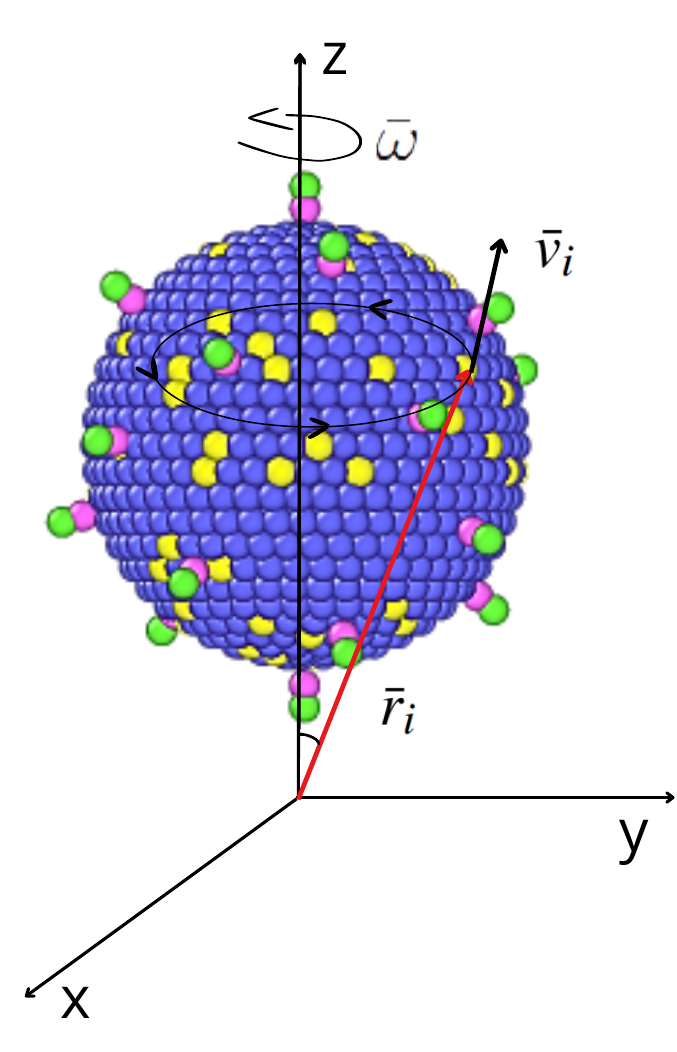}
  \caption{Schematic representation of the rigid virus model rotating around the $z$-axis with an angular frequency $\bar{\omega}$.}
  \label{fig3}
\end{figure}

In this work, the rotational dynamics of the virus model is quantified by the rotational diffusion coefficient $D_{r}$.
As described previously, the virus is treated as a rigid body composed of $N_v$ DPD particles, each with positions $\bar{r}_{1},\bar{r}_{2},...,\bar{r}_{N_v}$ and velocities $\bar{v}_{1},\bar{v}_{2},...,\bar{v}_{N_v}$ (see Fig. \ref{fig3}).
The instantaneous rotational velocity $\bar{\omega}$ of the whole virus model is obtained from the total angular momentum $\bar{L}$, which results from the contributions of all the constituent DPD particles \cite{Kim}:
\begin{equation}
\bar{L}=I \bar{\omega}=\sum_{n=1}^{N_{v}} \bar{r}_{n} \times m_{n} \bar{v}_{n}.
\label{momento}
\end{equation}

By definition, the components of the inertia tensor $I$ of the rigid rotating body about the axis of rotation are given by \cite{Goldstein},
\begin{equation}
I_{ik}=\sum_{n} m_{n} \left (  x_{nl}^{2} \delta_{ik} - x_{ni} x_{nk}\right ).
\label{tensor}
\end{equation}
Consequently, $\bar{\omega}$ can be determined as, 
\begin{equation}
    \bar{\omega}= I^{-1} \bar{L},
    \label{velo}
\end{equation}
where $I^{-1}$ denotes the inverse of the inertia tensor.
The inverse is computed in terms of the determinant of $I$ and its adjoint matrix,
\begin{equation}
    I^{-1}=\frac{Adj(I)}{\left | I \right |}.
    \label{inverse}
\end{equation}
Here, $Adj(I)$ denotes the adjoint matrix, defined as the transpose of the matrix of cofactors of $I$.
Once the angular velocity is calculated, the cumulative rotational displacement over the time interval $[t_{j_{0}},t_{k}]$ is computed as,
\begin{equation}
 \bar{\theta}(t_{k})=\sum_{j=j_{0}}^{k-1}\bar{\omega}(t_{j})\Delta \tau,   
\end{equation}
 where $\Delta \tau$ is the time step and $(\theta _{x}, \theta _{y}, \theta _{z })$ are the rotation components around the Cartesian axes. 
 The angular displacement between two frames separated by $m$ time steps is then \cite{Zhangg,Hunter,Han}
 \begin{equation}
    \Delta \bar{\theta}(t_{j};m)=\bar{\theta}(t_{j+m})-\bar{\theta}(t_{j}).
    \label{Deltatheta}
\end{equation}
Thus, from Eq. (\ref{Deltatheta}), the unbounded mean-square angular displacement (MSAD) can be expressed as \cite{Hunter,Spagnolo}:
\begin{equation}
\label{MSAD}
   \left \langle  \Delta \bar{\theta} ^{2} (t_{j};m) \right \rangle = \frac{1}{N_{t}-m} \sum_{j=j_{0}}^{N_{t}-m}\left|\left|\Delta \bar{\theta}(t_{j};m)  \right|\right|^{2},
 \end{equation}
where $N_{t}$ denotes the total number of frames considered to perform the corresponding average.

For sufficiently long times, MSAD grows linearly and follows a linear relationship with time \cite{Kammerer, Han, Hunter},
 \begin{equation}
   \lim_{t\rightarrow \infty } \left \langle   \Delta \bar{\theta} ^{2} \right \rangle =4D_{r}t,
   \label{Delta_Dr}
 \end{equation}
which allows the extraction of the rotational diffusion coefficient $D_{r}$ within the diffusive regime.

\begin{figure*}
\includegraphics[width=0.85\textwidth]{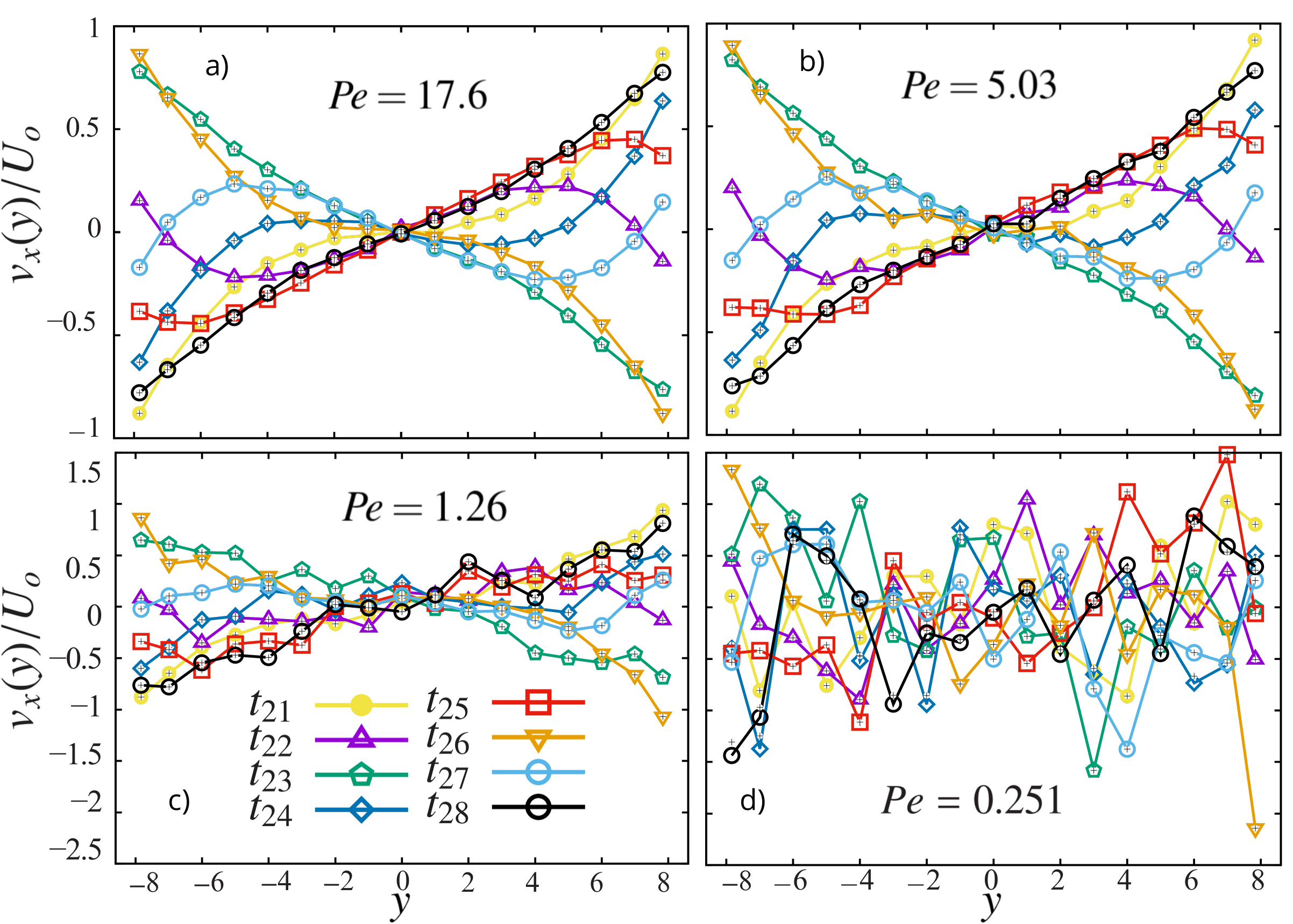}
\caption{\label{A}Velocity profiles of a confined DPD fluid subjected to oscillatory shear at a constant frequency, $\omega$, averaged over $10$ independent realizations. The velocity component $v_{x}(y)$ is normalized by the velocity $U_{o}$ (see Eq. (\ref{upper}) or Eq. (\ref{lower}). Profiles are recorded over a time interval ranging from $2.1$ to $\sim 2.8\times 10^{4}$ time steps, corresponding to frames from $t_{21}$ to $t_{28}$, respectively, where the subscripts denote frame numbers sampled every $1000$ time steps.
(a) $Pe = 17.6$: the oscillatory shear flow fully dominates over thermal fluctuations.
(b) $Pe = 5.03$: the shear flow remains dominant, although thermal motion introduces noticeable fluctuations.
(c) $Pe = 1.26$: thermal motion becomes dominant over the imposed shear, leading to significant profile distortions while a weak shear signature persists.
(d) $Pe = 0.251$: thermal fluctuations dominate, effectively suppressing the shear-induced velocity profile. }
\end{figure*}

\section{Hydrodynamic behavior of a confined aqueous host solvent: coupling between thermal fluctuations and imposed shear flow}
\subsection{General details of the simulation of the host fluid flow}  \label{GralSimDetails}
In the proposed model, the repulsion parameter $a_{ij}$ is specified according to the values listed in Table I. The noise amplitude, $\sigma = 3$, is chosen to satisfy the equilibrium condition \cite{Groot}, the temperature is fixed at $k_BT=1$, and the friction coefficient $\gamma$ can be obtained based on the fluctuation-dissipation theorem (see Eq. (\ref{termo})) \cite{Espanol}. 

All simulations presented here were performed using the parallel software package LAMMPS \cite{lammps}. 
The equations of motion were numerically integrated using the velocity-Verlet algorithm with a mid-step velocity update and a time step of $\Delta t = 0.03$ \cite{NIKUNEN03, lammps}. The unit of time is defined as $r_c\sqrt{m/k_B T}$, where $r_c$ is the unit of length (see Section \ref{DPD-II}), $k_B T$ is the thermal energy and is chosen as the unit of energy, and $m$ is the mass of each DPD particle \cite{EspanolWarren, Nguyen}, here considered as the unit of mass. 

As mentioned above, the model system consists of $56,700$ fluid particles, $49,340$ wall particles, $819$ particles that form the surface of the virus and two particles per peplomer. These particles are arranged to model a DPD fluid confined between two parallel, explicitly constructed planar walls separated by a distance $2h$. The lower wall is located at $y=-h$ and the upper wall at $y=h$, with $h=8.5$ (in units of $r_{c}$). As a result, the fluid is confined within a simulation volume of $30 \times 17 \times 30$ (in units of $r_{c}^{3}$). The results reported here are obtained by averaging over 10 independent realizations, which has led to a substantial reduction in the associated statistical uncertainties.


\subsection{Fluid response under oscillatory shear \label{Osci}}

We first report the response of the DPD host fluid under oscillatory shear, i.e., without virus. To this end, we have considered an oscillatory flow with constant frequency $\omega$ \cite{Wei, Allen}, and the rest of the parameters described in the previous paragraphs. 

 The walls oscillate at opposite velocities in the $x$ direction, with the following boundary conditions:
\begin{equation}
    v_{x}(h,t)=U_{0} cos (\omega t),
    \label{upper}
\end{equation}
\begin{equation}
    v_{x}(-h,t)=-U_{0}  cos (\omega t),
    \label{lower}
\end{equation}
where $U_{0}$ is the velocity amplitude. Since the confined fluid experiences a time-dependent rate of deformation, the shear rate can be calculated as $\kappa(y, t) = \partial v_{x}(y, t) / \partial y$. Accordingly, the expression for the shear rate at maximum velocity amplitude is given by
\begin{equation}
\kappa_{max}=\frac{U_{0}}{h}.
\label{max}
\end{equation}
Therefore, we use Eq. (\ref{max}), which represents the maximum shear rate, to define the Péclet number at constant frequency as,
\begin{equation}
Pe=\kappa_{max}r_{c}^{2}/D _{0}.
\label{Péclet}
\end{equation}

Fluid particles are initially randomly distributed between the confining walls. Once the simulation begins, the imposed oscillatory shear generates different flow regimes depending on the Péclet number ($Pe$). Then, for each value of $Pe$, the hydrodynamic field or velocity profile is fully determined and explicitly reported in Fig. \ref{A}. Our findings are summarized in the following paragraphs.

For $Pe \sim 17$, as shown in Fig. \ref{A} (a), the oscillatory shear flow dominates over thermal fluctuations due to the high shear rate, resulting in a well-defined laminar oscillatory velocity profile.
As the Péclet number decreases to $Pe \sim 5$ (Fig. \ref{A} (b)), hydrodynamic shear effects remain dominant, but thermal fluctuations begin to introduce slight deviations from the ideal oscillatory structure. The persistence of fluctuations around the averaged profile suggests that thermal effects remain relevant even in the presence of shear.

At $Pe \sim 1$ (Fig. \ref{A}(c)), the host fluid enters a transitional regime in which thermal fluctuations contribute substantially in addition to the imposed shear, consistent with the expectation that both effects have comparable magnitudes. Consequently, the flow field begins to weaken and exhibits slight irregularities in this regime.
For $Pe < 1$ (Fig. \ref{A}(d)), thermal motion dominates fully, and the velocity profile reflects the random dynamics characteristic of a thermally driven fluid.  

In summary, the sequence of hydrodynamic fields shown in Fig. \ref{A} revealed a distinct shift from a shear-controlled oscillatory regime to one dominated by thermal effects, underscoring how sensitively the behavior of the mesoscale fluid depends on the interplay between hydrodynamic forcing and thermal fluctuations. The $Pe$-dependence of the velocity profile will be crucial for understanding how the virus responds when exposed to an oscillatory shear field. This point will be discussed further below.

\section{Oscillatory shear effects on the rotational diffusion of a virus}
\begin{figure}[h]
\centering
  \includegraphics[height=6.53cm]{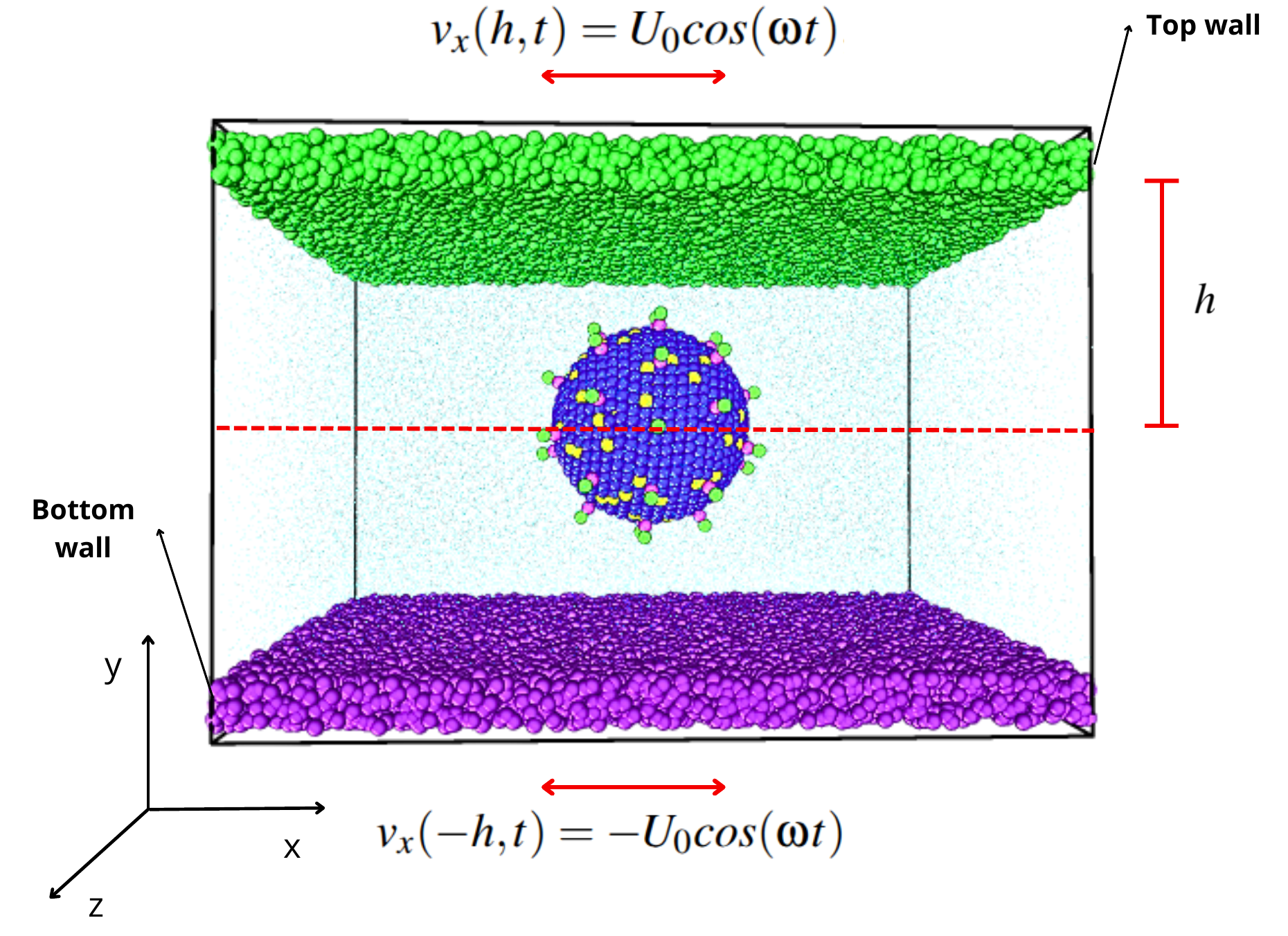}
  \caption{Initial configuration of the virus model subjected to oscillatory shear flow between two parallel walls. The red arrows indicate the oscillatory velocities applied to the walls. The wall separation is $2h$, and the virus is placed at the center of the simulation box, without contacting the walls. Fluid beads are rendered at a reduced size to improve the visibility of the virus.}
  \label{AB}
\end{figure}

We now turn to investigating how the virus behaves under oscillatory shear flow. The shear is applied to the solvent through rigid wall beads, and the virus responds to the hydrodynamic forces generated by this induced flow.

As discussed in the previous section, we characterized the hydrodynamic behavior of the host fluid in a wide range of oscillatory flow regimes determined by the interplay between oscillatory shear and thermal fluctuations and characterized by the Péclet number. Based on these results, we now examine how the virus behaves within these distinct hydrodynamic conditions. This hydrodynamic characterization enables us to assess the diffusivity of the virus in scenarios where multiple types of forces compete, offering a more realistic representation of the dynamics expected in complex biological systems, such as coronavirus suspensions.

To characterize the dynamical response of the virus, we determine its mean-square angular displacement (MSAD), $\left \langle \Delta \bar{\theta} ^{2} (t)  \right \rangle$ (see Eq. (\ref{MSAD})), from which the rotational diffusion coefficient is obtained by fitting the long-time behavior of the MSAD, as indicated in Eq. (\ref{Delta_Dr}).
 Fig. \ref{AB} shows a visualization of the virus model immersed in a DPD fluid subjected to oscillatory shear. The upper wall (green beads) oscillates with an angular frequency $\omega$ and an amplitude $U_0$. In contrast, the lower wall (purple beads) oscillates in the opposite direction with the same amplitude and frequency, following the same oscillatory flow setup described in Eqns. (\ref{upper}) and (\ref{lower}), respectively. 

To ensure that the whole system reaches steady state and improve statistical accuracy, the DPD equations are integrated over $10^{7}$ time steps, corresponding to a total simulation time of $t=(10^{7})(0.03~\tau) \sim  10^{5} \tau $, where $\tau = \sqrt{mr_c^2/k_BT}$. The positions and velocities of the particles are recorded every $10^3$ time steps. Using these trajectories, the MSAD is calculated up to $\sim 3 \times10^{4}~\tau$. Therefore, the rotational diffusion coefficient $D_{r}$ (in units of $\sqrt{k_{B}T/r_{c}^{2}m}$) for each value of the Péclet number, $Pe$, and virus model (specified by the number of peplomers, $N_{s}$) is obtained using Eq. (\ref{Delta_Dr}).


\begin{figure*}
\includegraphics[width=0.85\textwidth]{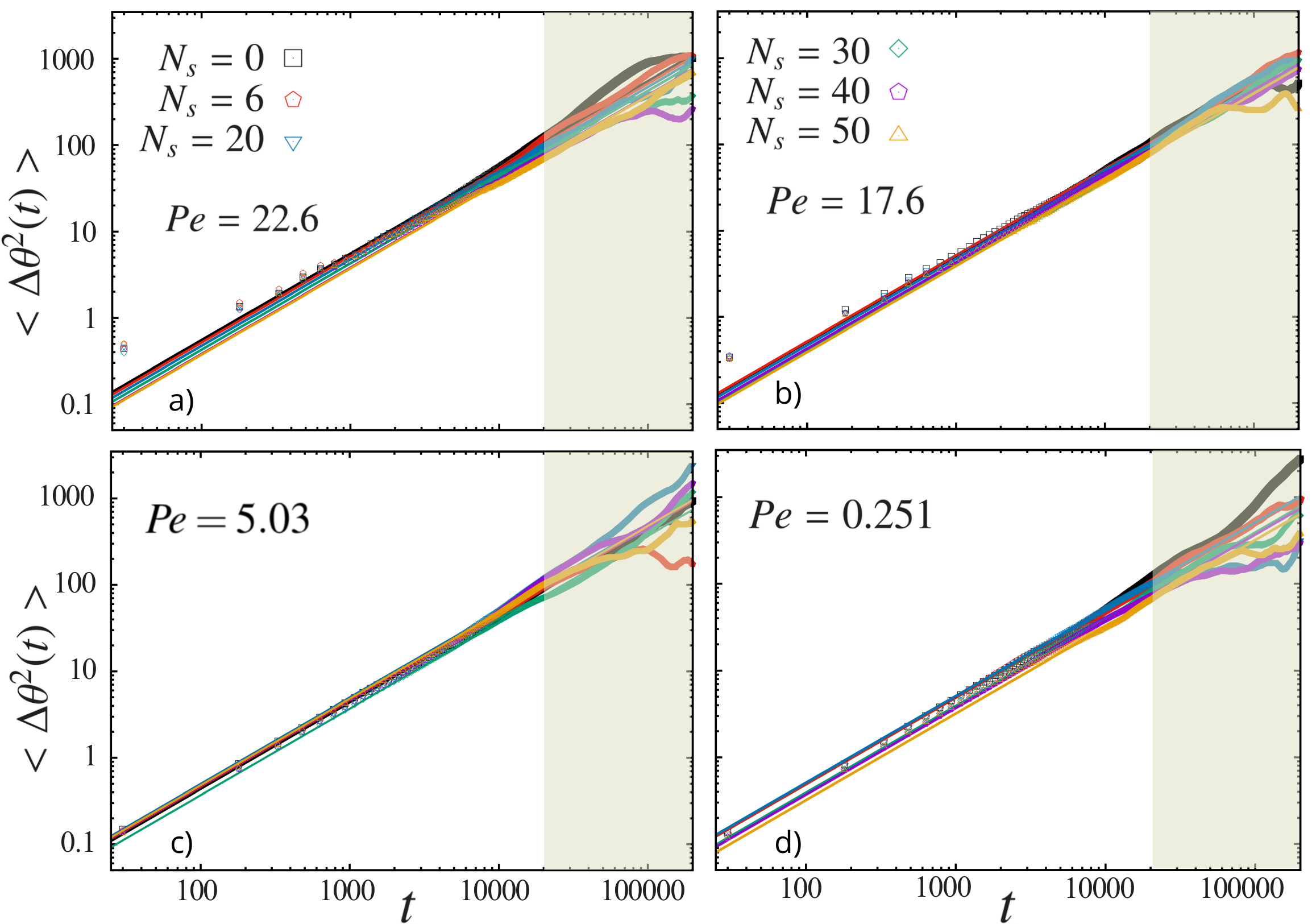}
\caption{\label{picos}Mean-square angular displacement (MSAD) of the virus model as a function of time ($t$) for four different Péclet numbers ($Pe$) for a specific value of the number of peplomers, $N_{s}$, as indicated. Each color corresponds to a different number of peplomers in the virus model. The shaded brown region indicates the noise-dominated long-time regime.
Symbols represent simulation data, and the solid lines correspond to linear fits in the diffusive regime.
MSADs for a) $Pe = 22.6$, b) $Pe = 17.6$, c) $Pe = 5.03$ and $Pe = 0.251$. Time is reported in simulation units.}
\end{figure*}

As discussed in the previous section, the value of $Pe$ is controlled by varying the shear amplitude. Fig. \ref{picos} shows four representative MSAD panels corresponding to distinct $Pe$ values: a) $Pe=22.6$, b) $Pe=17.6$, c) $Pe=5.03$, and d) $Pe=0.251$. In each panel, each curve corresponds to a virus model with a different number of peplomers, ranging from zero to fifty spikes.

For high $Pe$ values (Figs. \ref{picos} (a) and (b)), a first crossover appears around $t \sim 500$, followed by a second crossover at $t \sim 3000$, beyond which the MSAD reaches a long-time diffusive regime.  In contrast, at intermediate and low $Pe$ (see Figs. \ref{picos} (c) and (d)),  this early-time characteristic ($t \sim 500$) is not clearly observed,  and the system transitions more directly to diffusive behavior. The early-time crossover observed at high $Pe$ is attributed to transient advection associated with strong shear flow, since this crossover is not observed when thermal fluctuations dominate or compete. All curves exhibit a similar scaling consistent with diffusive in the time interval $\sim 10^3<t<10^4$, which is used for linear fits (solid lines). MSAD for $t>2\times10^4$ is not taking in count for the linear fit (brown shaded region on panels of Fig. \ref{picos}) since noise dominates at long-times.

For high and low $Pe$ (Figs. \ref{picos} (a), (b), and (d)), the rotational diffusion coefficient $D_r$ shows an overall decreasing trend as the number of peplomers increases. 
In contrast, at intermediate values of the Péclet number, $Pe$ (Fig. \ref{picos} (c)), this dependence becomes attenuated and less sharply defined, owing to the competing influences of shear and thermal fluctuations.
These observations suggest that the rotational dynamics of the virus is influenced by its surface architecture at high and low Péclet numbers, even though the peplomers are small compared to the viral diameter.

\begin{figure*}
\includegraphics[width=0.85\textwidth]{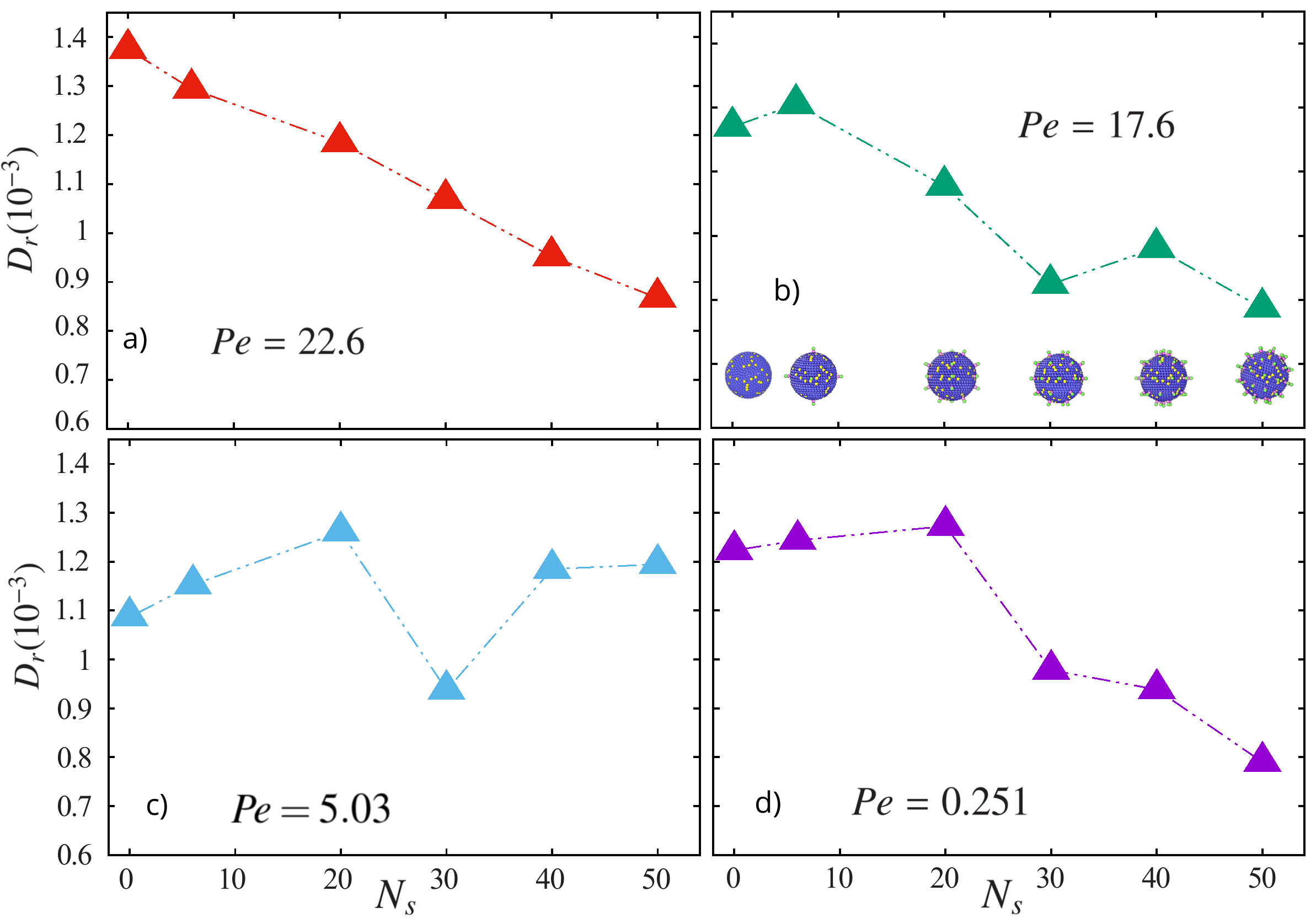}
\caption{\label{rotational}Rotational diffusion coefficient ($D_{r}$) of the virus model as a function of the number of peplomers ($N_{s}$) for four Péclet numbers ($Pe$), as indicated. a) For $Pe = 22.6$, $D_{r}$ tends to decrease with increasing $N_{s}$. b) For $Pe = 17.6$, a similar logarithmic trend is observed. c) For $Pe = 5.03$, $D_{r}$ no longer depends on $N_{s}$. d) For $Pe = 0.251$, $D_{r}$ decreases with $N_s$. Dashed lines in panels are included solely as visual guides to highlight the overall trend of $D_{r}$ and do not imply any physical significance.}
\end{figure*}
After analyzing the rotational dynamics of the virus model in Fig. \ref{picos}, the rotational diffusion coefficient $D_r$ was obtained by fitting the linear regime of the MSAD curves (see Eq. (\ref{Delta_Dr})).  Figure \ref{rotational} shows $D_r$ as a function of peplomer concentration,  organized into four panels corresponding to the same values of $Pe$ reported in Fig. \ref{picos}.

In Fig. \ref{rotational}(a) ($Pe = 22.6$), $D_r$ decreases from ($1.375 \pm 0.01$ to $0.867 \pm 0.01 $) $\times 10^{-3}$ as the number of peplomers increases. Error estimates correspond to the fitting uncertainties. A similar trend is observed for $Pe = 17.6$ (see Fig. \ref{rotational} (b), where the oscillatory shear forces remain dominant over thermal fluctuations; $D_r$ decreases from ($1.27 \pm 0.01$ to $0.99 \pm 0.02$) $\times 10^{-3}$. Thus, when oscillatory shear effects dominate over thermal fluctuations ($Pe \gtrsim 10$), an overall decreasing trend is observed: increasing the number of peplomers $N_s$ tends to reduce the rotational diffusion coefficient.

Interestingly, at high $Pe$ values (Figs. \ref{rotational}(a) and (b)), $D_r$ shows a decreasing trend with increasing $N_s$. This trend should be interpreted as an empirical observation within the finite range $6 \leq N_s \leq 50$, rather than as a universal scaling law.

Although an increase in the hydrodynamic radius could, in principle, affect rotational diffusion under equilibrium conditions, the observed decrease cannot be solely attributed to a simple geometric effect, since the peplomers are short and do not significantly modify the effective hydrodynamic size of the particle. Instead, the dependence of the effective rotational diffusion on the number of peplomers may arise from the characteristics of the hydrodynamic field (see Fig.  \ref{A}(a)), where the interplay between oscillatory forcing, confinement, and surface anisotropy becomes relevant. This interplay can modify the orientational dynamics by introducing additional flow-induced torques. In the high-Péclet-number regime, the oscillatory flow is expected to enhance the sensitivity of rotational motion to surface anisotropy, leading to an overall reduction of the effective rotational diffusion coefficient. In this regime, the dynamics is also influenced by the hydrodynamic torques generated by the oscillatory flow acting on the surface peplomers.

For $Pe < 10$, the relationship between $D_{r}$ and $N_{s}$, in particular, for $Pe = 5.03$, where the oscillatory shear competes directly with the thermal fluctuations, $D_r$ varies non-monotonically between ($1.08 \pm 0.01$ and $1.19 \pm 0.01$) $\times 10^{-3}$, i.e. without showing a clear correlation with $N_s$, as seen in Fig. \ref{rotational}(c). 
It is important to note that in this intermediate regime,  the contributions from oscillatory shear and thermal fluctuations are comparable, with shear being slightly dominant, while thermal fluctuations remain significant (see Fig. \ref{A}(b)).

For $Pe = 0.251$, thermal fluctuations dominate over oscillatory shear. As shown in Fig. \ref{rotational}(d), $D_r$ ranges between $(1.27  \pm 0.01 $ and $0.79 \pm 0.01)\times 10^{-3}$, with only a general correlation with $N_s$. In this regime, the rotational diffusion approaches its thermal limit, where the number of peplomers is expected to introduce only minor geometric corrections. However, in the present system, these corrections remain small, consistent with the short length of the peplomers, which only weakly perturb the hydrodynamic response. As a result, $D_r$ exhibits a decreasing trend with increasing $N_s$, which seems less pronounced than in the $Pe=22.6$ regime, and the possible amplification effects associated with oscillatory forcing are not observed.

Additionally, for the cases without peplomers ($N_s = 0$), the rotational diffusion deviates from the general trend. This deviation arises from the viral envelope model without dimers, which effectively imposes a slip boundary condition at the virus-solvent interface, enhancing the hydrodynamic slip and thus reducing rotational diffusion, as expected.

\begin{figure}[h]
\centering
  \includegraphics[height=6.15cm]{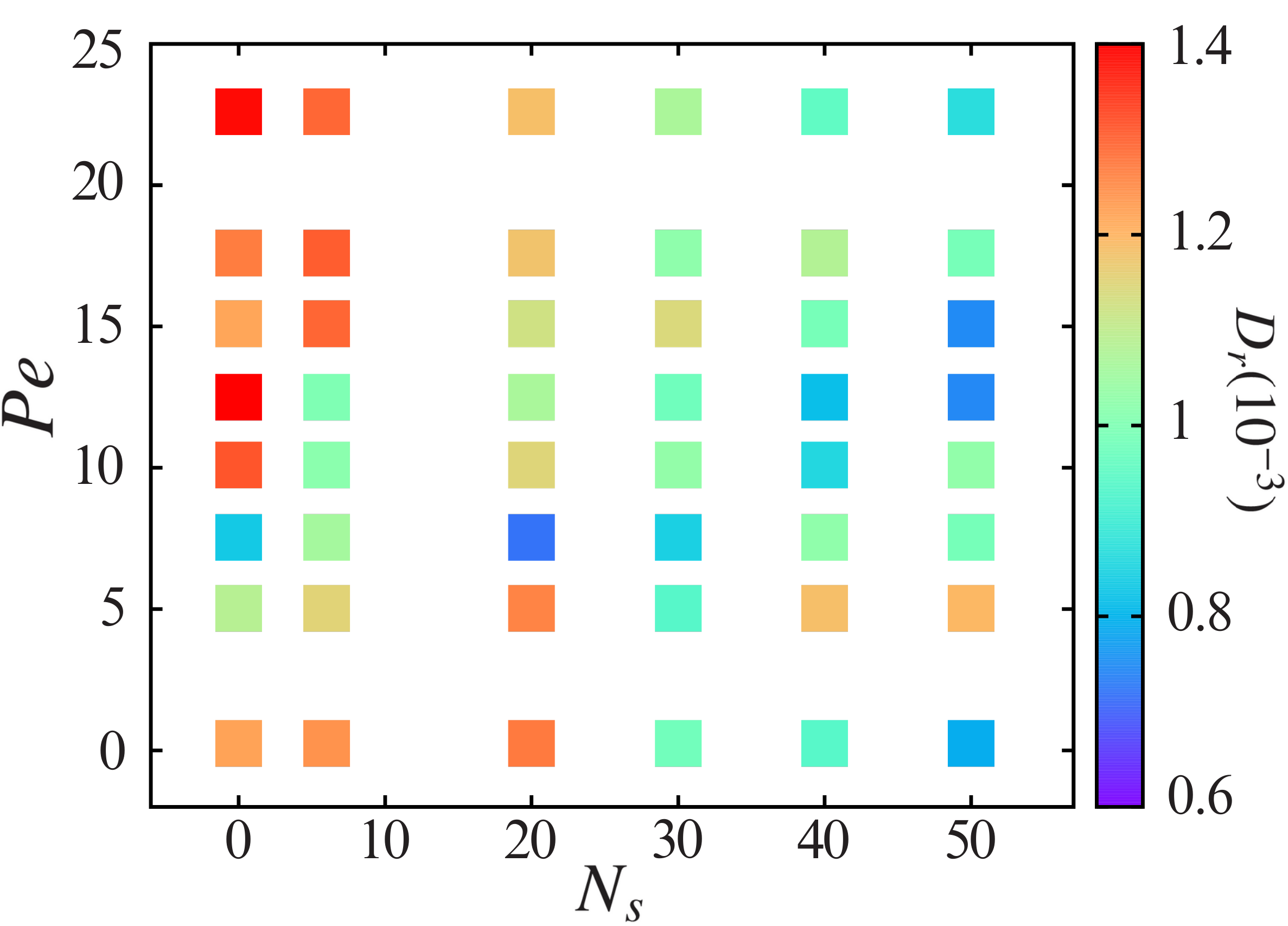}
  \caption{Color map of the rotational diffusion coefficient ($D_{r}$) of the virus model as a function of Péclet number ($Pe$) and number of peplomers ($N_{s}$). The color scale spans from blue for low values to red for high values.}
 \label{PhaseDiag}
\end{figure}

Fig. \ref{PhaseDiag} summarizes the rotational dynamics of the virus as a function of the transport regime and the number of peplomers. Although the peplomers are small compared with the viral diameter, they can still exert a measurable influence on the rotational dynamics in confined systems when oscillatory shear flow or thermal fluctuation dominate.

When oscillatory shear becomes dominant, advective transport is enhanced as the flow imposes a preferential direction on the motion, i.e., the virus follows a well-defined hydrodynamic field (see Fig. \ref{rotational}(a)). Consequently, for $Pe \gtrsim 10$, $D_r$ tends to fall within the higher-value region of the colormap for $N_s \leq 20$, indicating that rotational motion is more pronounced when the virus carries few or no peplomers. For $N_{s}  \geq  30$, $D_{r}$ falls within the blue region, showing that rotational diffusion decreases as the number of peplomers increases. 
In contrast, in the intermediate diffusive regime ($1\gtrsim Pe\gtrsim 10$), no clear pattern is observed in the color map of  $D_r$ shown in Fig. \ref{PhaseDiag}. This behavior is consistent with the flow fields displayed in Figs. \ref{A}(b) and \ref{A}(c),  which lack a well-defined structure and reflect the contributions of shear and thermal fluctuations. Finally, in the low- $Pe$ diffusive regime ($Pe=0.251$), where thermal fluctuations dominate, the pattern observed in the color map of  $D_r$ resembles that of the high- $Pe$  regime. This behavior can be understood from Fig. \ref{A}(d), where the hydrodynamic field is dominated by fluctuations and does not exhibit a preferential direction. This behavior at low $Pe$ has been further analyzed. To understand the role of confinement and the size of the peplomers, we performed simulations of the virus model embedded in a fluid with volume of dimensions $20$ $\times$ $20$ $\times$ $20$ ($r_c^3$). The number of peplomers was systematically varied and two peplomer sizes were considered: a dimer with a height of $1.5$$r_c$ and a pentamer with a height $2.5$ times larger than that of the dimer. For all simulations, the mean squared angular displacement (MSAD) was also calculated. Further details are provided in Figs \ref{appendixA} and \ref{appendixB} of the appendix \ref{Appendx}.

\begin{figure}[t]
\centering
  \includegraphics[height=6.3cm]{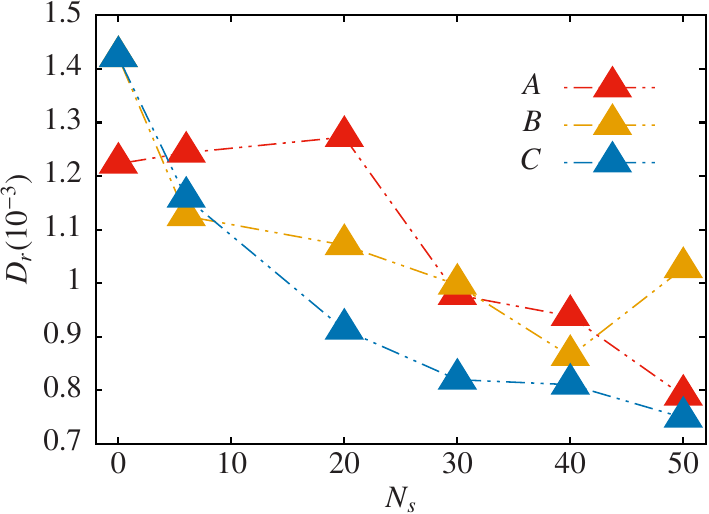}
  \caption{Rotational diffusion coefficient ($D_{r}$) of the virus model as a function of $N_{s}$, for three cases in which thermal motion dominates. $A$ (red circles) corresponds to the confined system at $Pe=0.251$. $B$ (orange squares) corresponds to an unconstrained (non-confined) system. $C$ (blue triangles) denotes a non-confined system with peplomers approximately $2.5$ times larger. } 
 \label{thermal-effect}
\end{figure}

In Fig. \ref{thermal-effect}, we compare the rotational diffusion coefficient as a function of the number of peplomers for three cases. The curve $A$ corresponds to $Pe=0.251$, where the diffusive regime dominates in a confined system; in this case, $D_r$ follows a general dependence on $N_{s}$ (red circles), same case as plot on Fig. \ref{rotational}(d).
In contrast, case $B$ represents the same virus model in a unbounded system, where a relationship between $D_{r}$ and $N_{s}$ emerges: $D_r$ clearly tends to decrease with the number of peplomers (orange squares). Moreover, when the virus is modeled with peplomers twice as large in a non-confined system, $D_r$ exhibits a clearly faster decrease with $N_s$ (blue triangles).
These results suggest the influence of the advection-diffusion competition for viruses suspended in confined fluids. In particular, a dependence between the rotational dynamics and the number of peplomers becomes more apparent when advective effects are dominant, i.e., at high $Pe$ values. In contrast, viruses suspended under bulk conditions exhibit a dependence between $D_r$ and $N_s$ even in the diffusive regime, where this dependence is also influenced by the size of the peplomers.

According to previous analytical results for viruses in bulk \cite{Kanso, Moreno}, the rotational diffusivity decreases monotonically with increasing peplomer concentration, in a regime where the dynamics is governed by thermal fluctuations. In contrast, for a virus confined in a restricted environment, a similar decreasing trend is observed at both high and low Péclet numbers. These results suggest that, under oscillatory forcing, the interplay between surface anisotropy and confinement contributes significantly to shaping the effective rotational dynamics of the virus.

\begin{figure}[h]
\centering
  \includegraphics[height=6.15cm]{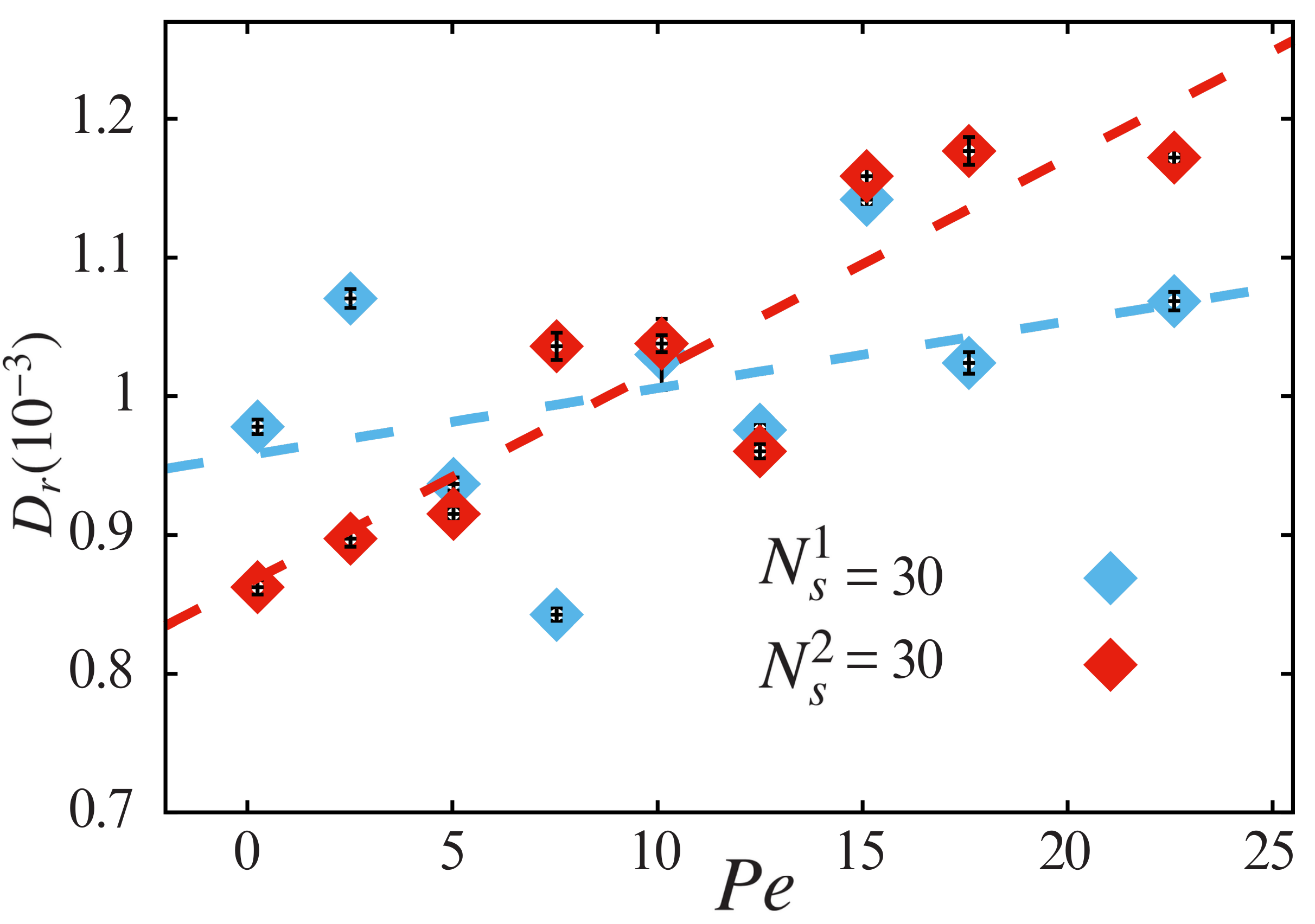}
  \caption{Rotational diffusion coefficient ($D_{r}$) of the virus model with $30$ peplomers as a function of the Péclet number. $N_{s}^{1}$ denotes the symmetric peplomer distribution (blue diamonds), while $N_{s}^{2}$ corresponds to the distribution experimentally reported by Yu et al. \cite{Yu} (red diamonds). The dashed lines are included as visual guides to emphasize the overall trend of $D_{r}$ only and do not imply any physical meaning.}
 \label{alpha}
\end{figure}

Additional to the confinement and peplomers size effect on the rotational diffussion, we have investigated how the spatial arrangement of spike proteins around the virus influences its rotational dynamics.
We have performed DPD simulations of our virus model with $N_s=30$ peplomers using two configurations, that is, a symmetric configuration where the spacing between peplomers is uniform, together with the experimentally observed distribution reported by Yu et al. \cite{Yu}. In both cases, $D_r$ increases with $Pe$, although the Yu-based configuration shows a steeper growth, as illustrated in Fig. \ref{alpha}. 
Interestingly, in the diffusive regime, the symmetric configuration exhibits a slightly higher $D_r$, while in the advective regime, $D_r$ is higher for $N_s^2$. This inversion can be explained by the fact that, at high $Pe$, the Yu-based virus model behaves effectively like one with fewer distinct peplomers, since some peplomers lie closer together; the contribution to the rotational dynamics can resemble that of a virus model with fewer peplomers when compared with the symmetric model. 
Nevertheless, the overall differences in $D_{r}$ between the two cases remain small, since the Yu-based virus model still maintains an approximately homogeneous distribution of peplomers on the viral surface.

The effect of the spike-protein distribution on viral diffusion was previously examined by Moreno et al. \cite{Moreno}. They reported the diffusivity of spike-decorated structures using mesoscopic hydrodynamic simulations. Their results showed that both the number and spatial arrangement of spikes, whether homogeneous or randomly distributed, significantly influence the diffusion of viruses such as SARS-CoV-2. In particular, random spike distributions break the symmetry of the virial envelope, whereas homogeneous distributions preserve it \cite{Moreno}.
Consistent with these findings, our results demonstrate that peplomer configurations also slightly affect viral rotational diffusion in a confined system. For virus models with $N_{s} = 30$, even small deviations from a symmetric peplomer arrangement led to some differences in diffusivity (see Fig. \ref{alpha}).

Consequently, we conclude that the effective rotational diffusion of the virus is controlled not only by its structural morphology or architecture, such as the number and spatial distribution of peplomers, but also by the dynamical properties of the surrounding fluid, including the flow regime, thermal fluctuations, and confinement. These results suggest that the rotational response of spike-decorated particles is influenced by the interplay between surface anisotropy and external forcing, particularly under nonequilibrium conditions. The DPD methodology is well-suited for capturing these effects, as it enables the systematic exploration of rotational dynamics across a wide range of flow regimes and environmental conditions.

\section{Conclusions and perspectives}

In conclusion, the DPD technique allowed us to characterize the fluid environment surrounding a minimal virus-inspired particle. Because our primary objective was to analyze the rotational diffusivity of the rigid spherical body decorated with dimers, a detailed understanding of the fluid conditions was essential, as the dynamical response of the particle depends on the properties of the host medium and the hydrodynamic field. Consequently, we characterized the hydrodynamic behavior of the host fluid by computing velocity profiles, which elucidate the role of thermal motion under different shear conditions.

By simulating a fluid confined between explicit solid boundaries subjected to oscillatory shear, we reproduced laminar flow regimes at large Péclet numbers ($Pe$) and delineated parameter ranges in which thermal fluctuations persist along or dominate over the externally imposed flow. Within these hydrodynamic conditions, we examined the rotational diffusion of the minimal virus model in regimes governed predominantly by thermal fluctuations or shear forces, capturing physical ingredients relevant to confined soft-matter systems inspired by biological contexts, where multiple mechanisms or driving forces coexist.

We found that the rotational diffusion coefficient exhibited a dependence on the number of peplomers $N_{s}$ at high $Pe$, consistent with a regime where oscillatory shear enhances the role of surface anisotropy under confinement. In this regime, the rotational diffusivity showed a generally decreasing trend with increasing peplomer number.
At intermediate $Pe$ values, $D_r$ shows no clear dependence on $N_s$, while at low $Pe$ a general dependence is recovered.
The absence of a correlation between $D_{r}$ and $N_{s}$ in regimes characterized by competing forces, in which the thermal contribution remains significant, was ascribed to hydrodynamic effects induced by confinement in the walls. This interpretation was further corroborated by comparative analyzes with unconfined systems. In addition, the results indicated that even small peplomers could influence the relationship between $D_r$ and $N_s$, although this effect becomes more pronounced for larger peplomers, leading to stronger dependence, in agreement with previous results obtained using alternative approaches \cite{Kanso, Moreno}.

Overall, our findings indicated that, in confined systems, variations in the number of peplomers influence rotational diffusion when oscillatory shear or thermal effects completely dominate. This finding is relevant to understanding how the rotational dynamics of virus-inspired particles can be modified in complex environments where multiple forces compete, such as those encountered by enveloped viruses in respiratory pathways \cite{Kaler}. Furthermore, the minimal virus model considered here may serve as a useful starting point to explore microscopic contributions to the rheological response of virus suspensions, since rotational behavior under oscillatory forcing reflects the interplay between viscous and elastic effects \cite{Kanso3, Roca, Jara}.

In summary, the rotational response of a virus-inspired particle model immersed in a fluid with competing forces might provide qualitative insight into physical mechanisms that influence the alignment of the peplomer near a target surface, a process that has been suggested to play a role in virus–cell interactions \cite{Kanso}. 

Last but not least, the results also suggested that the effect of the peplomer distribution is a characteristic that must be considered at high values of the $Pe$ number. Future work may explore how variations in the symmetry and size of surface protrusions, as well as the degree of confinement, further influence the rotational dynamics of virus-inspired particles.

\begin{acknowledgments}
K.G.F. gratefully acknowledges SECIHTI through Grant CVU:1047374 for financial support.
The authors acknowledge the project 124/2024 of the Convocatoria Institucional de Investigación Científica de la Universidad de Guanajuato. This work was possible thanks to the access to HPC-time granted by the following institutions: a) LANCAD and SECIHTI on the supercomputer Miztli at DGTIC UNAM, calls 2024 and 2025.  b) Laboratorio de Supercómputo del Bajío CIMAT through project "Supercómputo como motor de colaboraciones academia-industria en conjunto con el Instituto de Innovación, Ciencia y Emprendimiento para la Competitividad para el Estado de Guanajuato (IDEA GTO)". c) INKARI (High-Performance Computing Facilities located at the Astronomical Observatory of CHARACATO-UNSA). R.C.-P.\ acknowledges financial support from SECIHTI (Grant No.\ CBF2023-2024-3350).

\end{acknowledgments}


\appendix


\section{Rotational diffusion of a virus in the absence of external forces
\label{Appendx}}

The simulation model consists of a DPD fluid in a simulation box with periodic boundary conditions (without a wall boundary condition as in the previous case). The positions of the fluid particles are initially randomly generated within a domain of $20 \times 20 \times 20$ (in units of $r_{c}$), and their velocities follow a Gaussian distribution.  

For this system, we performed a systematic calculation of the MSAD curves and the rotational diffusion coefficient for the case without external forces, considering all virus models.

\begin{figure}[h]
\centering
  \includegraphics[height=5.5cm]{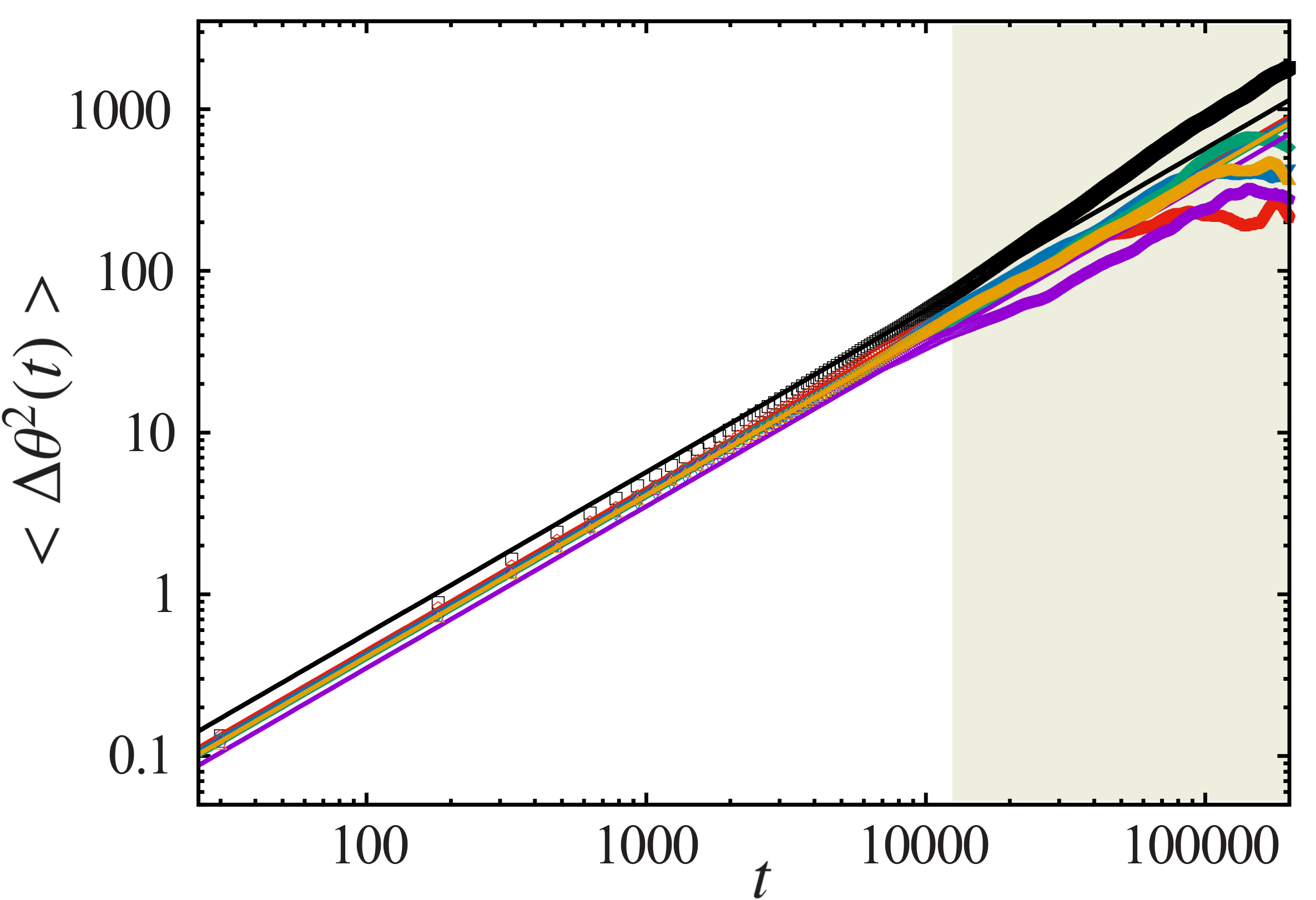}
  \caption{Mean-square angular displacement (MSAD) as a function of time ($t$). In this system, only thermal effects dominate. The dotted line is included solely as a visual guide to highlight the overall trend of $D_{r}$, without implying any physical significance. Reference virus model.}
 \label{appendixA}
\end{figure}

\begin{figure}[h]
\centering
  \includegraphics[height=5.5cm]{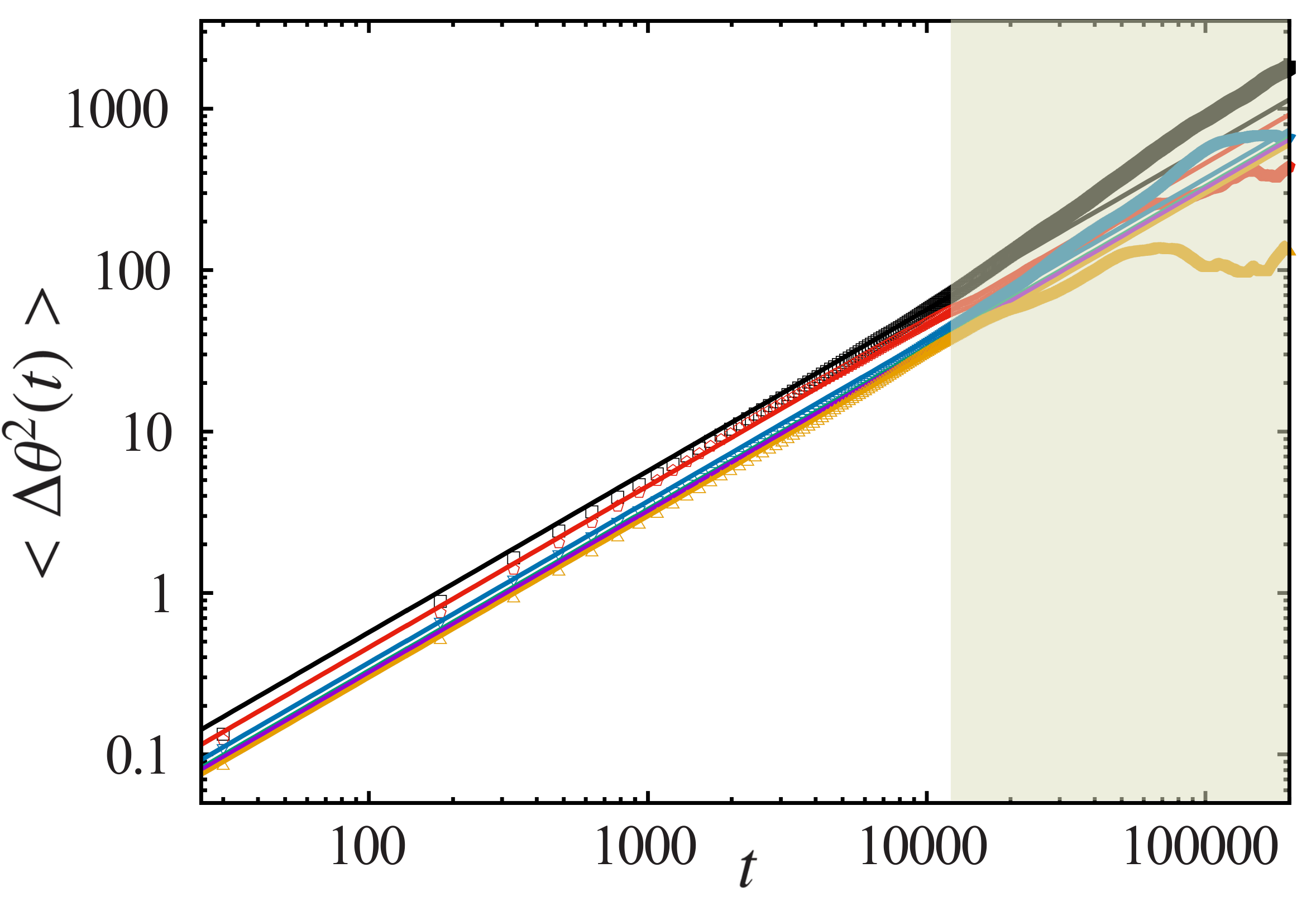}
  \caption{ Mean-square angular displacement (MSAD) as a function of time ($t$).n this system, only thermal effects dominate. Virus models featuring peplomers are approximately $2.5$ times larger than those in the reference model.}
 \label{appendixB}
\end{figure}




\nocite{*}
\bibliography{aipsamp}

\end{document}